\documentclass[sigconf]{acmart}

\usepackage{subcaption}

\renewcommand\footnotetextcopyrightpermission[1]{}
\usepackage{csvsimple}
\usepackage{amsthm}
\usepackage{stmaryrd}
\usepackage{qcircuit}

\usepackage{tikz}
\usetikzlibrary{arrows}
\usetikzlibrary{arrows.meta}
\usetikzlibrary{positioning}  
\usetikzlibrary{shadows}
\usetikzlibrary{backgrounds,calc,chains,
	matrix,positioning,shapes,shapes.geometric,decorations,
	shapes.arrows, decorations.pathmorphing, decorations.pathreplacing, decorations.markings}

\tikzset{
	font={\footnotesize},
	vertex/.style={draw,circle,inner sep=0pt,minimum width=0.5cm,minimum height=0.5cm},
	zeroterm/.style={below,inner sep=0pt,font=\tiny}
}

\newtheorem{defin}{Definition}
\newtheorem{myexample}{Example}

\newcommand{\ket}[1]{\ensuremath{\left|#1\right\rangle}}

\let\OLDthebibliography\thebibliography
\renewcommand\thebibliography[1]{
	\tiny
	\OLDthebibliography{\tiny #1}
}

\PassOptionsToPackage{hyphens}{url}
\usepackage{hyperref}

\begin{document}
	\sloppy
\title{Mapping Quantum Circuits to IBM QX Architectures\\Using the Minimal Number of SWAP and H Operations}
\date{}

\author{
	Robert Wille \hspace{1.5cm} Lukas Burgholzer \hspace{1.5cm} Alwin Zulehner \\
	{\normalsize Institute for Integrated Circuits, Johannes Kepler University Linz, Austria}\\
	{\normalsize robert.wille@jku.at \hspace{1.5cm} lukas.burgholzer@jku.at \hspace{1.5cm} alwin.zulehner@jku.at }
}

\begin{abstract}
	The recent progress in the physical realization of quantum computers (the first publicly available ones---IBM's QX architectures---have been launched in 2017) has motivated research on automatic methods that aid users in running quantum circuits on them. Here, certain physical constraints given by the architectures which restrict the allowed interactions of the involved qubits have to be satisfied. Thus far, this has been addressed by inserting SWAP and H operations. However, it remains unknown whether existing methods add a minimum number of SWAP and H operations or, if not, how far they are away from that minimum---an $\mathcal{NP}$-complete problem. In this work, we address this by formulating the mapping task as a symbolic optimization problem that is solved using reasoning engines like Boolean satisfiability solvers. By this, we do not only provide a method that maps quantum circuits to IBM's QX architectures with a \emph{minimal} number of SWAP and H operations, but also show by experimental evaluation that the number of operations added by IBM's heuristic solution exceeds the lower bound by more than 100\% on average. An implementation of the proposed methodology is publicly available
	at \url{http://iic.jku.at/eda/research/ibm_qx_mapping}.
\end{abstract}

\maketitle

\sloppy
\section{Introduction}
\label{sec:introduction}

Quantum computing~\cite{NC:2000} currently gains considerable momentum.
First concepts showing the (theoretical superiority) of this new computation paradigm
have already been developed a few decades ago (e.g.,~by Shor's famous factorization algorithm proposed in~\cite{Sho:94}
or Grover's database search iteration proposed in~\cite{Gro:96}). 
While this mainly electrified the academic community only, a real ``push'' of this topic 
was caused by the involvement of ``big players'' such as IBM, Google, Intel, or Microsoft in the recent years---additionally triggering the interest of industry as well as the public at large~\cite{gomes2018quantumcomputing,hsu2018quantumcomputing,courtland2017google}.
They are aiming to utilize quantum computers for applications such as quantum chemistry, optimization, machine learning, cryptography, quantum simulation, or solving systems of linear equations~\cite{preskill2018quantum}.

Here, particularly IBM's approach stands out, which provided the first publicly available quantum processor (in 2017) that can be accessed by everyone (not only academics) through cloud
access. Since then, their machines have been used by more than 100,000 users, who have run more than 6.5 million experiments thus 
far~\cite{ibmQ}.
This motivated research on automatic methods that aid users in running quantum circuits on the corresponding machines (known as \emph{IBM QX architectures}).

An obvious problem is thereby how to efficiently realize the desired quantum functionality on a respectively given  architecture. 
The functionality is usually represented by means of a \emph{quantum circuit} composed of qubits and quantum operations. \emph{Qubits} provide the 
basic entity of quantum computers which, as in conventional computation, may assume two basis states~$\ket{0}$ and $\ket{1}$ but additionally also any superpositions of them. 
\emph{Quantum operations} can be defined by arbitrary unitary matrices but have to be decomposed into elementary operations which are supported by the given architecture.
While for decomposing arbitrary quantum functionality to a sequence of elementary operations many solutions already exist  (by specifying the decomposition manually~\cite{cross2017open} or using automated approaches as described in~\cite{MWZ:2011,DBLP:journals/tcad/AmyMMR13}), further constraints need to be addressed.

In fact, logical qubits used in the originally given quantum circuit description cannot arbitrarily be mapped to physical qubits used in the QX architectures, but have to satisfy certain constraints defined by a 
coupling map.
This can be accomplished by adding \emph{SWAP} and \emph{H operations}---increasing the size of the circuit and, by this, harming the fidelity of the execution.
Accordingly, this raises the question of how to derive a proper mapping of logical qubits to physical qubits while, at the same time, minimizing the number of added SWAP and H operations---an $\mathcal{NP}$-complete problem as recently proven in~\cite{botea2018compiling}.
In the recent past, several solutions for this have been proposed~\cite{qiskit,zulehner2017efficient,zulehner2018compiling,siraichi2018qubit,li2018tackling}.
Moreover, even competitions seeking the best possible solution for this problem have been conducted in order to further trigger development in this area (see~\cite{ibmDeveloperChallenge}).

However, none of the methods presented thus far solves this problem in an \emph{exact}, i.e.,~a minimal, fashion. Instead, heuristics are applied. While this is a reasonable strategy for an $\mathcal{NP}$-complete problem, it leads to a significant uncertainty about the quality of the developments outlined above: Even if significant progress can be observed, it still remains completely unclear how far the proposed heuristics are from the optimum and, hence, how good the proposed methods really are? 
Because of this, \emph{exact} methods that can generate minimal (or at least close-to-minimal) results are essential for a substantial evaluation of the current \mbox{state of the art}---even if the optimum can only be generated for small instances.

In this work, we are proposing such solutions. 
To this end, we consider \emph{all} possible applications of the SWAP and H operations that may influence the realization of an originally given circuit on a QX architecture---a computationally very expensive task.
In order to cope with this complexity, we propose to utilize 
powerful reasoning engines such as solvers for Boolean satisfiability that can cope with large search spaces.
Besides that, additional performance optimizations are proposed that may lead to solutions which are not guaranteed to be minimal anymore, but can significantly speed up the solving time while
still generating at least close-to-minimal solutions. This is confirmed by experimental evaluations, which additionally show that IBM's heuristic mapping solution exceeds the lower bound by more than 100\% on average.

\section{Background}
\label{sec:background}

To keep this paper self contained, we briefly review quantum circuits as well as IBM's QX architectures in this section.

\subsection{Quantum Circuits}
\label{sec:quantum_circuits}

Quantum circuits are a frequently used description means for quantum computations. Here,  the logical qubits are represented by vertical circuit lines, while quantum gates describe (from left to right) the order in which quantum operations (whose functionality is described by unitary matrices) are applied to the qubits.
Common operations that act on a single qubit are $X$, $H$, and $T$, which negate the state of a qubit, set it into superposition, or apply a phase shift, respectively.\footnote{Since the actual functionality of the single quantum operations are not relevant in this work, we refer, e.g.,~to~\cite{NC:2000} for a more detailed treatment on that.}
In a quantum circuit diagram, these operations are denoted by square boxes labeled with a corresponding identifier.

Besides operations working on a single qubit, there also exist operations acting on multiple ones. A subset of such operations are controlled operations. Here, a single qubit operation is applied to the  target qubit if all controlling qubits are in the basis state~$\ket{1}$. \linebreak One commonly used representative is the CNOT gate, where a single control qubit determines whether the state of the target qubit is inverted or not.
The CNOT gate plays an important role in quantum computations, since it---together with single qubit operations---provides a universal set for quantum computing, i.e., any quantum computation can be decomposed into a sequence of CNOT and single qubit gates. 

In the remainder of this work, we are using the following notation for quantum circuits:

\begin{defin}
	\label{def:circuit}
	Let $Q = \{q_1, q_2, \ldots, q_j, \ldots, q_{n}\}$ be a set of $n$ logical qubits. Then, a quantum circuit $G = g_1g_2\cdots g_k\cdots g_{|G|}$ is a sequence of quantum gates, where each gate $g_k$ is either a single qubit gate $\mathbf{U}_k(q_j, \mathit{U})$ with target qubit $q_j \in Q$ and a unitary matrix $\mathit{U}$ describing the corresponding functionality, or a controlled not gate $CNOT_k(q_c, q_t)$ with control qubit $q_c \in Q$ and target qubit $q_t \in Q$ (with $q_c \neq q_t$).
\end{defin}

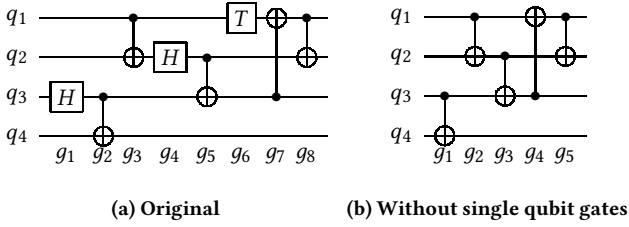
\begin{figure}
	\begin{subfigure}[b]{0.55\linewidth}
	\begin{center}	
	\mbox{\Qcircuit @C=.45em @R=0.4em @!R {
			\push{\rule{1em}{0em}} & \lstick{q_1} & \qw & \qw & \ctrl{1} & \qw & \qw & \gate{T} & \targ & \ctrl{1} & \qw \\
			\push{\rule{1em}{0em}} & \lstick{q_2} & \qw & \qw & \targ & \gate{H} & \ctrl{1} & \qw & \qw & \targ & \qw \\
			\push{\rule{1em}{0em}} & \lstick{q_3} & \gate{H} & \ctrl{1} & \qw & \qw & \targ & \qw & \ctrl{-2} & \qw & \qw \\
			\push{\rule{1em}{0em}} & \lstick{q_4} & \qw & \targ & \qw & \qw & \qw & \qw & \qw & \qw & \qw\\
			& & \ustick{g_1} & \ustick{g_2} & \ustick{g_3} & \ustick{g_4} & \ustick{g_5} & \ustick{g_6} & \ustick{g_7} & \ustick{g_8} \\				
	}}
\end{center}
	\caption{Original}
\label{fig:circuit}
	\end{subfigure}
	\begin{subfigure}[b]{0.44\linewidth}
	\begin{center}	
		\mbox{\Qcircuit @C=.45em @R=0.85em @!R {
				\push{\rule{1em}{0em}} & \lstick{q_1} & \qw & \ctrl{1} & \qw & \targ & \ctrl{1} & \qw \\
				\push{\rule{1em}{0em}} & \lstick{q_2} & \qw & \targ & \ctrl{1} & \qw & \targ & \qw \\
				\push{\rule{1em}{0em}} & \lstick{q_3} & \ctrl{1} & \qw & \targ & \ctrl{-2} & \qw & \qw \\
				\push{\rule{1em}{0em}} & \lstick{q_4} & \targ & \qw & \qw & \qw & \qw & \qw\\
				& & \ustick{g_1} & \ustick{g_2} & \ustick{g_3} & \ustick{g_4} & \ustick{g_5} \\				
		}}
	\end{center}
	\caption{Without single qubit gates}
\label{fig:circuit_trimmed}
\end{subfigure}

\caption{Quantum circuit to be mapped}
\end{figure}

\begin{myexample}
	Fig.~\ref{fig:circuit} shows a quantum circuit composed of 4 qubits and 8 gates. The single qubit gates $H$ and $T$ are visualized by square boxes labeled with $H$ and $T$, respectively, while the control and target qubit of CNOT gates are denoted by $\bullet$ and $\oplus$, respectively.
\end{myexample}

\subsection{IBM QX Architectures}
\label{sec:ibm_qx}

IBM's QX architectures have been made publicly available through cloud access in 2017 to allow conducting quantum computations on real devices. 
The IBM QX architectures provide the universal single qubit gate $U(\theta, \phi, \lambda) = R_z(\phi)R_y(\theta)R_z(\lambda)$ that is composed of two rotations around the $z$-axis and a rotation around the $y$-axis. Since the $U$ gate is universal, any single qubit operation can be conducted by specifying the parameters $\theta$, $\phi$, $\lambda$. By 
additionally supporting CNOT gates, IBM QX architectures allow for universal quantum computing (even though still limited in the number of qubits and gate fidelity).

To run quantum circuits on IBM's QX architectures, 
the respective logical qubits have to be mapped to \emph{physical} ones and gates have to be decomposed into a sequence of $U$ and $CNOT$ gates. In this work, we assume that the decomposition step has already been conducted (by specifying the decomposition manually~\cite{cross2017open} or using automated approaches as described in~\cite{MWZ:2011,DBLP:journals/tcad/AmyMMR13}). However, even then further constraints given by the architectures (called \mbox{\emph{CNOT-constraints}}) have to be satisfied. They state that not all physical qubits can interact with each other and, hence, CNOT gates cannot be applied to arbitrary pairs of physical qubits. Moreover, even if a CNOT gate can be applied to two physical qubits, it is restricted which physical qubit may serve as control and which physical qubit as target. These constraints are defined in a coupling map.

\begin{defin}
	\label{def:coupling_map}
  Let $P = \{p_1,p_2,\ldots, p_i, \ldots,p_m\}$ be a set of $m$ physical qubits available in the architecture. Then, the \emph{coupling map} \mbox{$CM \subseteq P \times P$} defines
   which physical qubits can interact with each other. More precisely, $\left( p_i, p_j\right) \in CM$ states that a CNOT gate with control qubit $p_i$ and target qubit $p_j$ can be applied, while
   \mbox{$\left( p_i, p_j\right) \notin CM$} states that a CNOT gate with control qubit $p_i$ and target qubit $p_j$ can \emph{not} be applied.
\end{defin}

\begin{figure}
	\centering
		\begin{tikzpicture}[terminal/.style={draw,rectangle,inner sep=2pt}]
\matrix[matrix of nodes,ampersand replacement=\&,every node/.style={vertex},column sep={1cm,between origins},row sep={0.75cm,between origins}] (qmdd) {
	\node (n5) {$p_5$}; \& \& \node (n1) {$p_1$}; \\
	\& \node (n3) {$p_3$}; \&  \\
	\node (n4) {$p_4$}; \& \& \node (n2) {$p_2$}; \\
};
\draw[-Implies,line width=0.5pt,double distance=4pt] (n2.north) -- (n1.south);
\draw[-Implies,line width=0.5pt,double distance=4pt] (n3.north east) -- (n1.south west);
\draw[-Implies,line width=0.5pt,double distance=4pt] (n3.south east) -- (n2.north west);
\draw[-Implies,line width=0.5pt,double distance=4pt] (n4.north east) -- (n3.south west);
\draw[-Implies,line width=0.5pt,double distance=4pt] (n4.north) -- (n5.south);
\draw[-Implies,line width=0.5pt,double distance=4pt] (n5.south east) -- (n3.north west);
\end{tikzpicture}

\caption{Coupling map of IBM QX4~\cite{ibmQX4}}
\label{fig:coupling_map}
\end{figure}
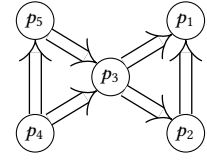

\begin{myexample}
Figure~\ref{fig:coupling_map} graphically represents the coupling map $CM = \{\left(p_2, p_1\right), \left(p_3, p_1\right), \left(p_3, p_2\right), \left(p_4, p_3\right), \left(p_4, p_5\right), \left(p_5, p_3\right)\}$ for the IBM QX4 architecture as directed graph. More precisely, the physical qubits $p_1, p_2,\ldots,p_5$ are visualized as vertices, whereas an entry $\left(p_i, p_j\right) \in CM$ is visualized as directed edge from $p_i$ to $p_j$. Hence, an arrow pointing from $p_i$ to $p_j$ indicates that a CNOT gate with control qubit $p_i$ and target qubit $p_j$ can be applied.
\end{myexample}

To satisfy the CNOT-constraints, one has to map the $n$ logical qubits $q_1,q_2,\ldots,q_n$ of the circuit to the $m\ge n$ physical qubits $p_1,p_2,\ldots,p_m$ of the considered quantum device such that all constraints given by the corresponding coupling map are satisfied. Unfortunately, it is usually not possible to find a mapping such that the constraints are satisfied throughout the whole circuit. More precisely, the following problems may occur:

\begin{itemize}
	\item A CNOT gate $CNOT_k(q_c, q_t)$ shall be applied while $q_c$ and $q_t$ are mapped to physical qubits $p_i$ and $p_j$, respectively, and $(p_i, p_j)\notin CM$ as well as $(p_j, p_i) \notin CM$.
	
	\item A CNOT gate $CNOT_k(q_c, q_t)$ shall be applied while $q_c$ and $q_t$ are mapped to physical qubits $p_i$ and $p_j$, respectively, and $(p_i, p_j)\notin CM$ while $(p_j, p_i) \in CM$
\end{itemize}

To overcome these problems, one strategy is to insert additional gates into the circuit to be mapped. 
More precisely, to overcome the first issue, one can insert SWAP operations into the circuit that exchange the state of two physical qubits and, by this, move around the logical ones 
(a strategy which also has been applied to make quantum circuits nearest neighbor-compliant~\cite{DBLP:journals/tcad/WilleLD14}).

\begin{figure}
		\begin{center}
		\mbox{
			\Qcircuit @C=.55em @R=0.45em {
				\push{\rule{4em}{0em}} & \lstick{p_1 \shortleftarrow q_1} & \qswap      & \qw & \push{q_2} & \push{\rule{.3em}{0em}\phantom{\equiv}\rule{.8em}{0em}} & \targ & \ctrl{1} & \targ & \qw & \push{\rule{.3em}{0em}\phantom{\equiv}\rule{.8em}{0em}} & \targ & \gate{H} & \targ & \gate{H} & \targ & \qw\\
				\push{\rule{4em}{0em}} & \lstick{p_2 \shortleftarrow q_2} & \qswap \qwx & \qw & \push{q_1} & \push{\rule[-5.5mm]{.3em}{0em}\equiv\rule{.8em}{0em}} & \ctrl{-1} & \targ & \ctrl{-1} & \qw & \push{\rule[-5.5mm]{.3em}{0em}\equiv\rule{.8em}{0em}} & \ctrl{-1} & \gate{H} & \ctrl{-1} & \gate{H} & \ctrl{-1} & \qw \\
		}}
	\end{center}	

	\caption{Decomposition of a SWAP operation}
	\label{fig:swap}
\end{figure}
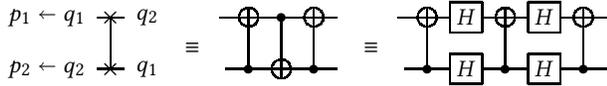

\begin{myexample}
	Fig.~\ref{fig:swap} shows the effect of a SWAP gate as well as its decomposition into elementary gates supported by the QX architectures. Assume that the logical qubits $q_1$ and $q_2$ are initially mapped to the physical ones $p_1$ and $p_2$, respectively (indicated by $\shortleftarrow$). Then, by applying a SWAP gate, the states of $p_1$ and $p_2$ are exchanged---eventually yielding a mapping where $q_1$ and $q_2$ are mapped to $p_2$ and~$p_1$, respectively.
\end{myexample} 

The second issue may also be solved by inserting SWAP operations. However, it is cheaper (fewer overhead is generated) to insert four Hadamard  operations (labeled by $H$) as they switch the direction of the CNOT gate (i.e.,~to change the target and the control qubit). This can also be observed in Fig.~\ref{fig:swap}, where $H$ gates switch the direction of the middle CNOT in order to satisfy all CNOT-constraints given by the coupling map.

As cost metric, we utilize the number of operations, since each operation introduces an error with a certain probability. Therefore, inserting a SWAP operation increases the cost by 7 (cf.~Fig.~\ref{fig:swap}), whereas switching the direction of a CNOT gate increases the cost by 4 (since 4 $H$ gates are added). Obviously, the general objective is to keep the overall cost as low as possible to  keep the overall fidelity as high as possible.

\section{Determining a Minimal Solution}
\label{sec:implementation}

Recently, first solutions have been proposed which derive a proper mapping of logical qubits to physical qubits while, at the same time, satisfying the CNOT constraints (see, e.g.,~\cite{qiskit,zulehner2017efficient,zulehner2018compiling,siraichi2018qubit,li2018tackling}).
However, as discussed in Section~\ref{sec:introduction}, it remains unknown how well they address the general objective reviewed in the previous section (i.e.,~how well they keep the costs caused by adding SWAP and H operations as small as possible). In this section, we describe how to determine a minimal solution
for this problem. To this end, we first describe the main idea of tackling the underlying complexity of the problem and, afterwards, introduce a symbolic formulation describing the problem in terms of a Boolean function. This is eventually used to solve the problem 
by applying powerful and efficient reasoning engines.

\subsection{Main Idea for Tackling the Complexity}
\label{sec:sat}

In this work, we aim for determining minimal or at least \linebreak \mbox{close-to-minimal} solutions for mapping quantum circuits to IBM QX architectures.\footnote{As discussed in Section~\ref{sec:ibm_qx}, we solve the mapping problem by inserting SWAP and/or H operations and assume that quantum circuits are already decomposed into elementary operations. Moreover, we do not consider pre- or post-mapping optimizations (as, e.g.,~proposed in~\cite{qiskit,zulehner2018compiling}) that may be applied before/after the mapping, but solely consider the actual mapping process in an exact fashion.}
To this end, we cannot heuristically or incompletely consider the search space but have to consider 
\emph{all} possible applications of the SWAP and H operations that may influence the realization of an originally given circuit on a QX architecture.
Obviously, this results in a computationally very expensive task ($\mathcal{NP}$-complete as recently proven in~\cite{botea2018compiling}). In order to cope with this complexity, we propose to utilize powerful reasoning engines such as solvers for Boolean satisfiability that can deal with large search spaces.
These solvers address the satisfiability problem defined as follows:

\begin{defin}\label{def:sat}
The {\em satisfiability problem} determines an assignment to the variables of a Boolean function \mbox{$\Phi: \{0,1\}^n \rightarrow \{0,1\}$} such that $\Phi$ evaluates to $1$ or proves that no such assignment exists. In an extended interpretation, additionally an objective function~$\mathcal{F}$ defined by \mbox{$\mathcal{F}(x_1,\dots,x_{n})= \sum_{i=1}^{n} w_i \dot{x}_i$} with $w_1,\dots,w_{n} \in \mathbb{Z}$ and $\dot{x} \in \{\overline{x}_i, x_i\}$ is provided. In this case, an assignment is to be determined which does not only satisfy~$\Phi$ but, at the same time, minimizes ~$\mathcal{F}$.
\end{defin}

\begin{figure*}
	
	\scalebox{0.85}{
	\begin{tikzpicture}

\node[minimum height = 2.5cm] (init){};
		
\node[draw, minimum height=2.5cm, right=1.65cm of init] (g1) {
		\begin{tabular}{l}
			if $z^1 = 0$: \\
		\mbox{
			\Qcircuit @C=.55em @R=0.85em {
				\push{\rule{3.7em}{0em}} & \lstick{q_3} & \ctrl{1} & \qw\\
				\push{\rule{3.7em}{0em}} & \lstick{q_4} & \targ & \qw \\
		}} \\[17pt]
				if $z^1 = 1$: \\
			\mbox{\Qcircuit @C=.55em @R=0.45em {
	\push{\rule{2em}{0em}} & \lstick{q_3} & \gate{H} & \targ & \gate{H} & \qw\\
	\push{\rule{2em}{0em}} & \lstick{q_4} & \gate{H} & \ctrl{-1} & \gate{H} & \qw \\
}} \\
		\end{tabular}
		
};

\node[above = 0cm of g1] (l1) {$g_1$};
		
\node[draw, minimum height=2.5cm, right = 0.25cm of g1] (g2) {
			{\LARGE $\pi \in \Pi$}
	};
		\node[above = 0cm of g2.south] (label2) {
					$y_\pi^2 = 1$
	};
		
\node[draw, minimum height=2.5cm, right = 1.65cm of g2] (g3) {
	\begin{tabular}{l}
	if $z^2 = 0$: \\
	\mbox{
		\Qcircuit @C=.55em @R=0.85em {
			\push{\rule{3.7em}{0em}} & \lstick{q_1} & \ctrl{1} & \qw\\
			\push{\rule{3.7em}{0em}} & \lstick{q_2} & \targ & \qw \\
	}} \\[17pt]
	if $z^2 = 1$: \\
	\mbox{\Qcircuit @C=.55em @R=0.45em {
			\push{\rule{2em}{0em}} & \lstick{q_1} & \gate{H} & \targ & \gate{H} & \qw\\
			\push{\rule{2em}{0em}} & \lstick{q_2} & \gate{H} & \ctrl{-1} & \gate{H} & \qw \\
	}} \\
	\end{tabular}
};

\node[above = 0cm of g3] (l3) {$g_2$};

\node[draw, minimum height=2.5cm, right = 0.25cm of g3] (g4) {
	{\LARGE $\pi \in \Pi$}
};
\node[above = 0cm of g4.south] (label3) {
	$y_\pi^3 = 1$
};

\node[draw, minimum height=2.5cm, right = 1.65cm of g4] (g5) {
	\begin{tabular}{l}
	if $z^3 = 0$: \\
	\mbox{
		\Qcircuit @C=.55em @R=0.85em {
			\push{\rule{3.7em}{0em}} & \lstick{q_2} & \ctrl{1} & \qw\\
			\push{\rule{3.7em}{0em}} & \lstick{q_3} & \targ & \qw \\
	}} \\[17pt]
	if $z^3 = 1$: \\
	\mbox{\Qcircuit @C=.55em @R=0.45em {
			\push{\rule{2em}{0em}} & \lstick{q_2} & \gate{H} & \targ & \gate{H} & \qw\\
			\push{\rule{2em}{0em}} & \lstick{q_3} & \gate{H} & \ctrl{-1} & \gate{H} & \qw \\
	}} \\
	\end{tabular}
};

\node[above = 0cm of g5] (l3) {$g_3$};

		\draw ($(init.north east)!0.1!(init.south east)$) node[left] {$p_1$} -- ($(g1.north west)!0.1!(g1.south west)$) node[above, midway, yshift=-0.3em] {$x^{1}_{11}x^{1}_{12}x^{1}_{13}x^{1}_{14}$};
		\draw ($(init.north east)!0.3!(init.south east)$) node[left] {$p_2$} -- ($(g1.north west)!0.3!(g1.south west)$) node[above, midway, yshift=-0.3em] {$x^{1}_{21}x^{1}_{22}x^{1}_{23}x^{1}_{24}$};
		\draw ($(init.north east)!0.5!(init.south east)$) node[left] {$p_3$} -- ($(g1.north west)!0.5!(g1.south west)$) node[above, midway, yshift=-0.3em] {$x^{1}_{31}x^{1}_{32}x^{1}_{33}x^{1}_{34}$};
		\draw ($(init.north east)!0.7!(init.south east)$) node[left] {$p_4$} -- ($(g1.north west)!0.7!(g1.south west)$) node[above, midway, yshift=-0.3em] {$x^{1}_{41}x^{1}_{42}x^{1}_{43}x^{1}_{44}$};
		\draw ($(init.north east)!0.9!(init.south east)$) node[left] {$p_5$} -- ($(g1.north west)!0.9!(g1.south west)$) node[above, midway, yshift=-0.3em] {$x^{1}_{51}x^{1}_{52}x^{1}_{53}x^{1}_{54}$};
	
		\draw ($(g1.north east)!0.1!(g1.south east)$) -- ($(g2.north west)!0.1!(g2.south west)$);
		\draw ($(g1.north east)!0.3!(g1.south east)$) -- ($(g2.north west)!0.3!(g2.south west)$);
		\draw ($(g1.north east)!0.5!(g1.south east)$) -- ($(g2.north west)!0.5!(g2.south west)$);
		\draw ($(g1.north east)!0.7!(g1.south east)$) -- ($(g2.north west)!0.7!(g2.south west)$);
		\draw ($(g1.north east)!0.9!(g1.south east)$) -- ($(g2.north west)!0.9!(g2.south west)$);

		\draw ($(g2.north east)!0.1!(g2.south east)$) -- ($(g3.north west)!0.1!(g3.south west)$) node[above, midway, yshift=-0.3em] {$x^{2}_{11}x^{2}_{12}x^{2}_{13}x^{2}_{14}$};
		\draw ($(g2.north east)!0.3!(g2.south east)$) -- ($(g3.north west)!0.3!(g3.south west)$) node[above, midway, yshift=-0.3em] {$x^{2}_{21}x^{2}_{22}x^{2}_{23}x^{2}_{24}$};
		\draw ($(g2.north east)!0.5!(g2.south east)$) -- ($(g3.north west)!0.5!(g3.south west)$) node[above, midway, yshift=-0.3em] {$x^{2}_{31}x^{2}_{32}x^{2}_{33}x^{2}_{34}$};
		\draw ($(g2.north east)!0.7!(g2.south east)$) -- ($(g3.north west)!0.7!(g3.south west)$) node[above, midway, yshift=-0.3em] {$x^{2}_{41}x^{2}_{42}x^{2}_{43}x^{2}_{44}$};
		\draw ($(g2.north east)!0.9!(g2.south east)$) -- ($(g3.north west)!0.9!(g3.south west)$) node[above, midway, yshift=-0.3em] {$x^{2}_{51}x^{2}_{52}x^{2}_{53}x^{2}_{54}$};

		\draw ($(g3.north east)!0.1!(g3.south east)$) -- ($(g4.north west)!0.1!(g4.south west)$);
		\draw ($(g3.north east)!0.3!(g3.south east)$) -- ($(g4.north west)!0.3!(g4.south west)$);
		\draw ($(g3.north east)!0.5!(g3.south east)$) -- ($(g4.north west)!0.5!(g4.south west)$);
		\draw ($(g3.north east)!0.7!(g3.south east)$) -- ($(g4.north west)!0.7!(g4.south west)$);
		\draw ($(g3.north east)!0.9!(g3.south east)$) -- ($(g4.north west)!0.9!(g4.south west)$);

		\draw ($(g4.north east)!0.1!(g4.south east)$) -- ($(g5.north west)!0.1!(g5.south west)$) node[above, midway, yshift=-0.3em] {$x^{3}_{11}x^{3}_{12}x^{3}_{13}x^{3}_{14}$};
		\draw ($(g4.north east)!0.3!(g4.south east)$) -- ($(g5.north west)!0.3!(g5.south west)$) node[above, midway, yshift=-0.3em] {$x^{3}_{21}x^{3}_{22}x^{3}_{23}x^{3}_{24}$};
		\draw ($(g4.north east)!0.5!(g4.south east)$) -- ($(g5.north west)!0.5!(g5.south west)$) node[above, midway, yshift=-0.3em] {$x^{3}_{31}x^{3}_{32}x^{3}_{33}x^{3}_{34}$};
		\draw ($(g4.north east)!0.7!(g4.south east)$) -- ($(g5.north west)!0.7!(g5.south west)$) node[above, midway, yshift=-0.3em] {$x^{3}_{41}x^{3}_{42}x^{3}_{43}x^{3}_{44}$};
		\draw ($(g4.north east)!0.9!(g4.south east)$) -- ($(g5.north west)!0.9!(g5.south west)$) node[above, midway, yshift=-0.3em] {$x^{3}_{51}x^{3}_{52}x^{3}_{53}x^{3}_{54}$};

		\draw ($(g5.north east)!0.1!(g5.south east)$) -- ++ (0.5, 0) node[right]{$\dots$};
		\draw ($(g5.north east)!0.3!(g5.south east)$) -- ++ (0.5, 0) node[right]{$\dots$};
		\draw ($(g5.north east)!0.5!(g5.south east)$) -- ++ (0.5, 0) node[right]{$\dots$};
		\draw ($(g5.north east)!0.7!(g5.south east)$) -- ++ (0.5, 0) node[right]{$\dots$};
		\draw ($(g5.north east)!0.9!(g5.south east)$) -- ++ (0.5, 0) node[right]{$\dots$};

	\end{tikzpicture}
}

	\caption{Symbolic formulation for mapping the circuit shown in Fig.~\ref{fig:circuit}}
	\label{fig:formulation}
\end{figure*}
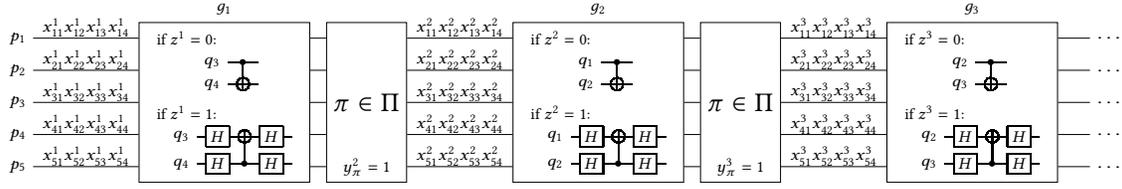

\begin{myexample}
	Let $\Phi = (x_1 + x_2 + \overline{x}_3) (\overline{x}_1 + x_3) (\overline{x}_2 +
	x_3)$. Then, $x_1=1$, $x_2=0$, and $x_3 = 1$ is a satisfying assignment solving the SAT problem. Additionally, let $\mathcal{F}=x_1+x_2+x_3$.
	Then, $x_1=0, x_2=0$, and $x_3 = 0$ is a solution which does not only satisfies~$\Phi$ but also minimizes~$\mathcal{F}$.
\end{myexample}

In the past, very efficient reasoning engines for the satisfiability problem have been proposed (see, e.g.,~\cite{WFG+:2007,de2008z3}). 
Instead of simply traversing the complete space of assignments, intelligent decision heuristics, powerful learning schemes, and efficient implication methods are applied and, by this, instances composed of numerous variables and constraints---defining huge search spaces---can be solved efficiently.
In this work, we are utilizing this reasoning power to consider the whole search space and, by this, to tackle the complexity.
To this end, however, a symbolic formulation is required which completely describes the problem and is provided in terms of a Boolean function (so that it can be used by those reasoning engines).

\subsection{Symbolic Formulation of the Problem}
\label{sec:core}

In the following, a symbolic formulation of the considered problem---mapping quantum circuits to IBM QX architectures using SWAP and H operations---is proposed. 
To this end, variables are defined which describe all possible applications of the SWAP and H operations. 
More precisely, those operations basically affect how logical qubits from an originally given quantum circuit are mapped to the physical qubits of an IBM QX architecture. The mapping might be changed before each gate.
Since only CNOT gates may cause violations of CNOT constraints,
we ignore single qubit gates when formulating the mapping problem.\footnote{Note that this additionally reduces the overall complexity of the problem to be solved.} This leads to the following symbolic formulation:

\begin{defin}
Let $G=g_1g_2\cdots g_k \cdots g_{|G|}$ be a quantum circuit composed of $|G|$ CNOT gates. Each gate $g_k$ operates on a (logical) control qubit $q_c$ and a (logical) target qubit $q_t$ (cf.~Def.~\ref{def:circuit}). Furthermore, let $Q = \{q_1,\hdots,q_j,\hdots,q_{n}\}$ be a set of $n$ logical qubits that shall be mapped to the $m\ge n$ physical qubits in \mbox{$P=\{p_1,\ldots, p_i, \ldots,p_{m}\}$}. Finally, let $CM \subseteq P \times P$ be the description of the coupling map indicating what circuit lines can interact with each other on the given architecture and how (cf.~Def.~\ref{def:coupling_map}). 
Then, \emph{mapping variables} $x^k_{ij}$ with $k\in\{1,\hdots,|G|\}$, $i\in\{1,\hdots,m\}$, and $j\in\{1,\hdots,n\}$ are introduced representing whether, before gate $g_k \in G$, the logical qubit $q_j$ is mapped to the physical qubit $p_i$ ($x^k_{ij}=1$) or not ($x^k_{ij} =0$). 
\end{defin}

\begin{myexample}
Consider again the circuit shown in Fig.~\ref{fig:circuit} (and assume the single qubit gates have been removed as shown in Fig.~\ref{fig:circuit_trimmed}). 
Then, Fig.~\ref{fig:formulation} sketches a symbolic formulation for mapping the circuit to IBM QX4 which represents all possible mappings of logical qubits to physical qubits.
For example, the leftmost part of Fig.~\ref{fig:formulation} represents the initial mapping of the logical qubits to the physical ones. 
Here, e.g.,~setting $x^1_{13}=1$ represents that logical qubit~$q_3$ is mapped to physical qubit~$p_1$ right before gate $g_1$. 
\end{myexample}

Passing this symbolic formulation to a reasoning engine would yield arbitrary assignments that most likely encode\linebreak impossible/useless mappings (e.g., mapping several logical qubits to the same physical one). Hence, we have to restrict the assignment of variables so that only valid solutions are obtained. To this end, we have to ensure that:

\begin{enumerate}
\item A well-defined mapping between logical and physical qubits is conducted (i.e.,~each logical qubit is uniquely assigned to \emph{exactly one} physical qubit and vice versa). This is ensured by 

\begin{equation} 
\bigwedge_{k=1}^{|G|} \left( \bigwedge_{j=1}^{n} \left(\sum_{i=1}^m x^k_{ij} = 1 \right) \wedge \bigwedge_{i=1}^{m}\left( \sum_{j=1}^n x^k_{ij} \leq 1\right) \right).
\end{equation} 

\item All gates only act on physical qubits that have a corresponding entry in the coupling map of the considered architecture or an entry where control and target qubit are switched.\footnote{Note that we additionally consider entries where the control and target qubit are switched since this can be handled by inserting $H$ gates. This way, the reasoning engines gets to decide what combination of SWAP and~$H$ gates yield the cheapest global mapping.} 
This is ensured by 
\begin{equation} 
\bigwedge_{g_k = CNOT_k(q_c,q_t)\in G} \left(\bigvee_{(p_i,p_j)\in CM} (x^k_{ic} \wedge x^k_{jt}) \vee (x^k_{it} \wedge x^k_{jc}) \right).
\end{equation} 	
\end{enumerate}

Adding these restrictions and passing the resulting symbolic formulation to a reasoning engine eventually yields a valid solution. Moreover, the resulting formulation covers the entire search space in a symbolic fashion.
Having this, all that is left is a proper description of the costs of the respectively chosen mapping. 
As reviewed in Section~\ref{sec:ibm_qx}, these costs accumulate from (1)~the costs for changing the mapping of the logical qubits to the physical ones throughout the circuit (by inserting SWAP operations) and (2)~the costs for switching the control and the target qubit for CNOT gates (by inserting H operations). 
The former one (changes on the mapping) may be applied before each CNOT gate (except the first one, which defines the initial mapping and, hence, can be set arbitrarily anyway); the latter one (switching the control/target qubits) may happen in each gate. To properly describe this within the symbolic formulation, we further introduce the following variables:

\begin{defin}
	Let $1 \le k \le |G|$ be the index of gate $g_k$ in a quantum circuit, $m$ the number of physical qubits in the considered quantum device, and $\pi \in \Pi$ a permutation of $m$ elements that indicates how the state of the physical qubits is permuted (eventually realized by inserting SWAP operations). Then, the \emph{permutation variables}~$y^k_\pi$ indicate whether the permutation $\pi$ is applied before gate $g_k$ ($y^k_\pi = 1$) or not ($y^k_\pi = 0$).
	Furthermore, the \emph{switching variables}~$z^k$ indicate whether the direction of the CNOT gate $g_k$ is switched ($z^k=1$) or not ($z^k=0$).
\end{defin}

\begin{myexample}
Consider again the symbolic formulation shown in Fig.~\ref{fig:formulation}.
The spots in the circuit where the mapping may change are sketched by boxes labeled $\pi$. Here, the variable assignment before and after (i.e.,~the assignments of $x^{k-1}_{ij}$ and  $x^k_{ij}$) may change according to a permutation~$\pi\in \Pi$ (eventually to be represented by $y^k_\pi$). Furthermore, in each gate~$g_k$, the $z^k$-variables define whether the direction of the CNOT gate is switched or not.
\end{myexample}

Using these variables, we can describe what permutation $\pi\in\Pi$ is applied to the states of the physical qubits of the circuit lines before each gate $g_k\in G$ by introducing
\begin{equation}\label{eq:permutations}
\bigwedge_{k=2}^{|G|} \left( \bigwedge_{\pi\in\Pi} \left( \bigwedge_{i=1}^{m}\bigwedge_{j=1}^{n} \left( x^{k-1}_{ij} = x^k_{\pi(i)j}\right)\right) \Leftrightarrow y^k_\pi\right).
\end{equation} 	

In fact, this ensures that $y^k_\pi$ is set to~1 iff the assignment of the variables $x^{k-1}_{ij}$ and  $x^k_{ij}$ indeed describe a change of the mapping defined by~$\pi$.\footnote{If $n<m-1$, $\pi$ cannot be determined uniquely. Then, a left-handed implication is required instead of an equivalence in Eq.~(\ref{eq:permutations})---in conjunction with a constraint that only one variable $y_\pi^k$ is assigned to 1. For sake of clarity of Eq.~(\ref{eq:permutations}), we assume that $n=m$.}

Similarly, we can describe for each CNOT gate~$g_k$ whether the direction of the control and target qubits are switched by introducing
\begin{equation} 	\label{eq:switches}
\bigwedge_{g_k = CNOT_k(q_c,q_t)\in G} \left(\bigvee_{(p_i,p_j)\in CM} (x^k_{it} \wedge x^k_{jc}) \right) \Leftrightarrow z^k.
\end{equation}
In fact, this ensures that $z^k$ is set to~1 iff, for gate~$g_k$, the control qubit is set to position~$j$ and the target qubit is set to position~$i$ although, according to the coupling map, it has to be vice versa (i.e.,~control and target qubits are switched).

Satisfying all of the  constraints introduced above yields a valid mapping of the originally given circuit to the desired architecture while, at the same time, the costs are determined by
\begin{equation} 	
\mathcal{F}= \sum_{k=2}^{|G|} \sum_{\pi\in\Pi} (7 \cdot \text{swaps}(\pi)\, y^k_\pi) + \sum_{k=1}^{|G|} (4 \cdot z^k).
\end{equation} 
Here, $\text{swaps}(\pi)$ defines the number of SWAP operations needed to realize the permutation~$\pi$. This has to be determined for each permutation $\pi\in\Pi$---a process, which needs to be conducted only once and can be done, e.g.,~by using an exhaustive search for the architectures considered in this work. 
By this, whenever the reasoning engine chooses a mapping which eventually creates a permutation~$\pi$ before gate~$g_k$, Eq.~(\ref{eq:permutations}) sets~$y^k_\pi=1$ and, hence, adds the corresponding costs (7 gates for each SWAP operation; cf.~Section~\ref{sec:background}) to the overall costs~$\mathcal{F}$.
Similarly, whenever the reasoning engine chooses a mapping  which requires switching the direction of a CNOT gate~$g_k$,  
Eq.~(\ref{eq:switches}) sets~$z^k=1$ and, hence, adds the corresponding costs (4~H operations; cf.~Section~\ref{sec:background})
to the overall costs~$\mathcal{F}$.

\subsection{Minimizing the Cost}

Passing the eventually resulting symbolic formulation to a reasoning engine allows to determine a valid mapping together with the associated cost (i.e.,~the number of additionally required elementary operations).
Since we are also interested in the minimum costs, the cost function~$\mathcal{F}$ needs to be further restricted. One direct solution could be to simply set~$\mathcal{F}$ to a fixed value and approach towards the minimum, e.g.,~by applying a binary search.
However, since many reasoning engines additionally allow to consider an objective function (cf.~Def.~\ref{def:sat}), the most efficient way is to simply 
add the objective~$\min: \mathcal{F}$ to the resulting instance---enforcing the reasoning engine not only to determine a satisfying assignment (representing a valid mapping) but, at the same time, also to minimize~$\mathcal{F}$.

\begin{myexample}\label{ex:solutions}
Passing the symbolic formulation sketched in~Fig.~\ref{fig:formulation} together with all constraints and the objective function to a reasoning engine, eventually yields a mapping (and a corresponding addition of SWAP and H operations) as shown in Fig.~\ref{fig:solution}.
This circuit provides a realization of the originally given circuit from Fig.~\ref{fig:circuit} which is applicable for the IBM QX architecture specified by the coupling map shown in Fig.~\ref{fig:coupling_map} and, at the same time, yields \emph{minimum} costs caused by additionally required SWAP and H operations ($\mathcal{F} = 4$).
\end{myexample}

\begin{figure}

\scalebox{1}{
		\mbox{\Qcircuit @C=.45em @R=0.4em @!R {
			\push{\rule{1em}{0em}} & \lstick{p_1 \shortleftarrow q_4} & \qw & \targ & \qw & \qw & \qw & \qw & \qw & \qw & \qw & \qw & \qw \\
			\push{\rule{1em}{0em}} & \lstick{p_2 \phantom{~\shortleftarrow q_4}} & \qw & \qw & \qw & \qw & \qw & \qw & \qw & \qw & \qw & \qw & \qw \\
			\push{\rule{1em}{0em}} & \lstick{p_3 \shortleftarrow q_3} & \gate{H} & \ctrl{-2} & \qw & \qw & \targ & \qw & \gate{H}&\ctrl{1}&\gate{H} & \qw & \qw \\
			\push{\rule{1em}{0em}} & \lstick{p_4 \shortleftarrow q_1} & \qw & \qw & \ctrl{1} & \qw & \qw & \gate{T} & \gate{H} & \targ& \gate{H} & \ctrl{1} & \qw\\
			\push{\rule{1em}{0em}} & \lstick{p_5 \shortleftarrow q_2} & \qw & \qw & \targ & \gate{H} & \ctrl{-2} & \qw & \qw & \qw & \qw& \targ & \qw \\
			&  & & \ustick{g_1} & \ustick{g_2} & & \ustick{g_3} & & & \ustick{g_4} & & \ustick{g_5} \\				
	}}
}
	\caption{Resulting circuit (with minimal SWAP/H costs) }
	\label{fig:solution}
\end{figure}

\section{Performance Improvements}
\label{sec:improvements}

Determining minimal solutions obviously is the desired way to go. However, even with powerful reasoning engines, we cannot always escape the $\mathcal{NP}$-complete complexity of the problem. In this regard, the methodology proposed in the previous section allows for several performance improvements.
In fact, the reasoning engine only needs to determine a ``minimal'' assignment for the $x^k_{ij}$-variables (the variables $y_\pi^k$ and $z^k$ can be ignored, since they are only used for formulating the costs and their assignments can directly be deduced from the $x^k_{ij}$-variables). 
For a quantum circuit composed of $n$ logical qubits and $|G|$ CNOT gates to an architecture with $m$ physical qubits, this leads to a total of 
$n\cdot m\cdot |G|$ variables to be assigned and, hence, an overall search space of $2^{n\cdot m \cdot |G|}$ which can easily be restricted by adding further constraints to the \mbox{$x^k_{ij}$-variables}. 
While this may lead to solutions which are not guaranteed to be minimal anymore, it can significantly speed up the solving time while, at the same time, remaining very close-to-minimal (as also confirmed by experimental evaluations summarized in Section~\ref{sec:results}).
This section shows possible improvements in this regard.

\subsection{Considering Subsets of Physical Qubits}
\label{sec:subset_qubits}

A scenario frequently occurs where the number~$n$ of logical qubits of a given quantum circuit to be mapped is smaller than the number~$m$ of physical qubits provided by the architecture (i.e.,~where $n < m$). Then, obviously, not all physical qubits are required. In fact, this allows to consider only a subset of $n$ physical qubits while ignoring the remaining $m-n$ ones.
Since the number of physical qubits to consider contributes to the search space in an exponential fashion, restricting this number yields substantial simplifications. In order to remain as close as possible to the minimal solution, one can try out all $\binom{m}{n}$ possible subsets of qubits to consider and solve the respectively resulting (smaller) instances separately. This reduces the overall search space to $\binom{m}{n} 2^{n^2 \cdot |G|}$. 

\begin{myexample}
	Consider again the symbolic instance for mapping a four qubit quantum circuit to a five qubit architecture sketched in Fig.~\ref{fig:formulation}. By considering only four physical qubits in the mapping procedure, the overall search space for a single instance reduces from $2^{4\cdot 5 \cdot 5} = 2^{100}$ to $2^{4^2 \cdot 5} = 2^{80}$. Even if all $\binom{5}{4} = 5$ possible subsets of physical qubits are considered separately, this still yields a significant reduction of the overall search space.
\end{myexample}

The search space can be reduced further by checking whether some of the physical qubits in a subset are isolated from others (this can be done in $O(n)$ time). 
If so, the instance for this subset does not have to be be passed to the reasoning engine as no solution can be found anyway.

\begin{myexample}
	Assume that the circuit over four qubits considered thus far shall be mapped to the IBM QX4 architecture shown in Fig.~\ref{fig:coupling_map}.
	Then, all subsets of physical qubits that have to be checked should contain $p_3$, since no connected sub-graph composed of four nodes without~$p_3$ is possible. This reduces the number of instances that are passed to the reasoning engine from $\binom{5}{4} = 5$ to $\binom{4}{3} = 4$. 
\end{myexample}

\subsection{Restricting the Possible Permutations}
\label{sec:limiting_permutations}

Thus far, we allowed permutations $\pi\in \Pi$ of the mapping before each gate (except the first one where an arbitrary initialization can be chosen).
While this guarantees minimality (since all possible solutions are considered), this substantially contributes to the complexity. 
In many cases, however, valid and cheap mappings are still possible if permutations of mappings are allowed not before \emph{all} gates~$g\in G$,
but only before a subset $G' \subseteq G\setminus\{g_1\}$ of them. 
With~$|G'|$ being significantly smaller than~$G$, this reduces the overall search space to $2^{n\cdot m\cdot (|G'|+1)}$. 
While applying this idea, $G'$ can be chosen arbitrarily. A smaller~$G'$ leads to a larger performance improvement, but also a more restricted instance (yielding solutions that might be far from minimal or even instances for which no valid mapping can be determined anymore).
In this work, the following strategies for defining~$G'$ are considered:\footnote{Many more strategies have been considered and evaluated, but are omitted in this paper due to space limitations.}

\begin{itemize}
\item \emph{Disjoint qubits}, which exploits the fact that gates acting on disjoint sets of qubits can always be mapped in a way that no intermediate permutations are required.\footnote{Note that such a set of gates is called \emph{layer} in some heuristic solutions~\cite{qiskit,zulehner2017efficient}.}
To this end, the quantum circuit is clustered into sequences of gates acting on disjoint sets of qubits and permutations are only allowed before each of those sequences.

\item \emph{Odd gates}, which allows permutations only before gates with an odd index (except for $g_1$). Here, it is still guaranteed that a valid mapping can be determined since either (1)~the gates operate on the disjoint sets of qubits as discussed above, (2)~the gates share both qubits, or (3)~the gates share one qubit (and there exists at least one qubit that can interact with two other qubits).

\item \emph{Qubit triangle}, which exploits the structure of architectures whose coupling map forms ``triangles'' of physical qubits as in case of, e.g.,~$p_1$, $p_2$, and $p_3$ in Fig.~\ref{fig:coupling_map}. Here, we can cluster the circuit into sequences of gates where each sequence acts on at most three qubits. Then, each such sequence of gates can be mapped to a triangle as described above and permutations are only required before each of those sequences.
\end{itemize}

\begin{myexample}
Consider the quantum circuit shown in Fig.~\ref{fig:circuit_trimmed}. 
Applying the strategies proposed above yields the following subsets~$G'$:
\begin{itemize}
\item \emph{Disjoint qubits}: $G' = \{g_3, g_4, g_5\}$, since the gates $g_1$ and $g_2$ operate on disjoint qubits (saving the permutation between $g_1$ and $g_2$).

\item \emph{Odd gates}: $G' = \{g_3, g_5\}$, since they constitute the odd gates in the circuit.
	
\item \emph{Qubit triangle}: $G' = \{g_2\}$, since all gates $g_2$, $g_3$, $g_4$, and $g_5$ operate on only three qubits and, hence, can be mapped to one of the ``triangles'' of the architecture without the need for further permutations. That is, only a permutation prior to~$g_2$ needs to be considered.
\end{itemize}
As can be seen, these strategies yield much more restrictive applications of permutations. While this substantially increases the performance of the solving process, it 
does not harm minimality in this case (but may for other circuits as evaluated in the next section). \end{myexample}

\section{Experimental Results}
\label{sec:results}

\begin{table*}
	\caption{Experimental results}
	\label{tab:results}
	
	\scriptsize
	\begin{tabular}{l|c|r||r|r||r|r||r|r|r||r|r|r||r|r|r||r}
		\multicolumn{3}{c||}{} & \multicolumn{2}{c||}{Min. (Sec.~\ref{sec:implementation})} & \multicolumn{2}{c||}{Perf. Opt. (Sec.~\ref{sec:subset_qubits})} & \multicolumn{9}{c||}{Performance Optimized (Section~\ref{sec:limiting_permutations})} & IBM~\cite{qiskit} \\	
		\multicolumn{3}{c||}{} & \multicolumn{2}{c||}{} & \multicolumn{2}{c||}{} & \multicolumn{3}{c||}{Disjoint qubits} & \multicolumn{3}{c||}{Odd gates} & \multicolumn{3}{c||}{Qubit triangle} \\
		Benchmark & $n$ & original cost & $c_{min}$ & $t\,[s]$ & $c~(\Delta_{\mathit{min}})$ & $t\,[s]$ & $|G'|$ & $c~(\Delta_{\mathit{min}})$ & $t\,[s]$ & $|G'|$ & $c~(\Delta_{\mathit{min}})$ & $t\,[s]$ & $|G'|$ & $c~(\Delta_{\mathit{min}})$ & $t\,[s]$ & $c~(\Delta_{\mathit{min}})$ \\\hline
		\begin{filecontents*}{results.csv}
name,n,gunary,gcnots,gall,costMin,tMin,costRed,tRed,diffRed,L1,costL1,tL1,diffL1,L2,costL2,tL2,diffL2,L3,costL3,tL3,diffL3,IBM,diffIBM
3\_17\_13,3,19,17,36,59,29,59,0,0,17,59,0,0,9,60,0,1,1,60,0,1,80,21
ex-1\_166,3,10,9,19,31,5,31,0,0,9,31,0,0,5,31,0,0,1,31,0,0,39,8
ham3\_102,3,9,11,20,36,10,36,0,0,11,36,0,0,6,36,0,0,1,36,0,0,48,12
miller\_11,3,27,23,50,82,231,82,0,0,23,82,0,0,12,82,0,0,1,82,0,0,82,0
4gt11\_84,4,9,9,18,34,7,34,0,0,9,34,0,0,5,34,0,0,2,34,0,0,37,3
rd32-v0\_66,4,18,16,34,63,281,63,35,0,16,63,35,0,8,63,1,0,2,72,0,9,101,38
rd32-v1\_68,4,20,16,36,65,276,65,35,0,16,65,36,0,8,65,1,0,2,74,0,9,99,34
4gt11\_82,5,9,18,27,62,133,62,137,0,18,62,139,0,9,62,3,0,5,62,1,0,77,15
4gt11\_83,5,9,14,23,49,17,49,17,0,14,49,18,0,7,50,1,1,3,50,0,1,65,16
4gt13\_92,5,36,30,66,109,528,109,533,0,29,109,199,0,15,110,10,1,9,110,5,1,126,17
4mod5-v0\_19,5,19,16,35,64,256,64,264,0,16,64,255,0,8,68,2,4,3,69,0,5,109,45
4mod5-v0\_20,5,10,10,20,35,10,35,10,0,10,35,11,0,5,35,0,0,3,35,0,0,64,29
4mod5-v1\_22,5,10,11,21,40,7,40,7,0,10,40,9,0,6,40,0,0,3,43,0,3,52,12
4mod5-v1\_24,5,20,16,36,63,54,63,55,0,16,63,56,0,8,63,3,0,3,63,0,0,98,35
alu-v0\_27,5,19,17,36,63,74,63,73,0,16,63,38,0,9,63,2,0,3,67,0,4,101,38
alu-v1\_28,5,19,18,37,64,94,64,92,0,17,64,44,0,9,67,10,3,3,68,0,4,123,59
alu-v1\_29,5,20,17,37,64,351,64,355,0,16,64,119,0,9,64,3,0,3,68,0,4,104,40
alu-v2\_33,5,20,17,37,64,42,64,44,0,17,64,44,0,9,64,4,0,4,64,0,0,99,35
alu-v3\_34,5,28,24,52,90,719,90,727,0,24,90,724,0,12,91,10,1,4,91,0,1,178,88
alu-v3\_35,5,19,18,37,64,103,64,101,0,17,64,74,0,9,64,3,0,3,68,0,4,121,57
alu-v4\_37,5,19,18,37,64,119,64,121,0,17,64,43,0,9,64,6,0,3,68,0,4,110,46
mod5d1\_63,5,9,13,22,48,14,48,13,0,11,48,8,0,7,48,5,0,5,48,1,0,98,50
mod5mils\_65,5,19,16,35,64,96,64,98,0,16,64,94,0,8,65,1,1,3,65,0,1,108,44
qe\_qft\_4,5,44,27,71,94,136,94,135,0,19,94,21,0,14,94,9,0,16,94,12,0,115,21
qe\_qft\_5,5,69,38,107,135,401,135,395,0,26,135,21,0,19,139,107,4,24,145,48,10,163,28
\end{filecontents*}
\csvreader[late after line=\\, late after last line=\\,]{results.csv}
		{1=\name,2=\qubits, 3=\uGates, 4=\CNOTs, 5=\allGates, 6=\costMin, 7=\tMin, 8=\costRed, 9=\tRed, 10=\diffRed, 11=\LA, 12=\costLA, 13=\tLA, 14=\diffLA, 15=\LB, 16=\costLB, 17=\tLB, 18=\diffLB, 19=\LC, 20=\costLC, 21=\tLC, 22=\diffLC, 23=\costIBM, 24=\diffIBM}
		{\name & \qubits & $\uGates + \CNOTs = \allGates$ & \costMin & \tMin & \costRed~(+\diffRed) & \tRed & \LA & \costLA~(+\diffLA) & \tLA & \LB & \costLB~(+\diffLB) & \tLB & \LC & \costLC~(+\diffLC) & \tLC & \costIBM~(+\diffIBM)}
	\end{tabular}
	~\\
	\raggedright{$n$: number of logical qubits \hspace*{0.5cm} \emph{original cost}: number of single qubit gates plus number of CNOT gates before mapping \hspace*{0.5cm} $c$: cost (number of operations) of the mapped circuit \hspace*{0.5cm} \\ $\Delta_{min}$: difference to minimum cost \hspace*{0.5cm} $t$: runtime in seconds}
	\vspace*{-3mm}
\end{table*}

The proposed method for mapping a quantum circuit to the IBM QX architectures using the minimal number of SWAP and H operations 
has been implemented in C++ (the implementation is publicly available at \url{http://iic.jku.at/eda/research/ibm_qx_mapping}). As reasoning engine, the \emph{Z3} solver~\cite{de2008z3} has been utilized. 
Afterwards, we conducted extensive evaluations using quantum circuits (also considered in previous work and taken from~\cite{WGT+:2008,cross2017open})
to be mapped to the IBM QX4 architecture~\cite{ibmQX4}. All evaluations have been conducted on an Intel Core i7-3930K machine with 4~GHz and 64~GB of main memory running Ubuntu~18.04.

The left part of Table~\ref{tab:results} thereby provides a selection of the results
obtained by a first series of evaluations---aiming for evaluating the effect of the performance improvements discussed in Section~\ref{sec:improvements}. More precisely, the first three columns describe the name of the considered quantum circuit, the number of logical qubits $n$, and the \emph{original cost} of the circuit (i.e.,~the number of single qubit gates plus the number of CNOT gates before mapping). 
In the remaining columns, we list the cost $c$ (i.e.,~the number of gates) of the obtained circuit and the runtime $t$ (in CPU seconds) required by Z3 when applying the method guaranteeing minimality discussed in Section~\ref{sec:implementation} as well as the adapted methods additionally incorporating the performance improvements discussed in Section~\ref{sec:improvements}. For the adapted versions, we additionally list the difference to the minimum (i.e.,~$\Delta_{min}$) in parenthesis. Moreover, for each strategy that limits the number of permutations, we additionally list  in columns denoted $|G'|$ how many permutations are allowed.

As can be seen by these results, determining a mapping with the minimum number of SWAP and H gates is quite expensive---only solutions for instances with rather few CNOT gates can be determined (which is not surprising since the underlying problem is $\mathcal{NP}$-complete). When considering only a subset of the physical qubits (cf.~Section~\ref{sec:subset_qubits}), we observe a significant reduction of the runtime for benchmarks with 3 or 4 qubits, while still preserving minimality. Limiting the number of permutation also has a tremendous effect on the runtime. In fact, the runtime required to solve an instance indirectly correlates  with $|G'|$. However, limiting the number of permutations too much generates rather poor results regarding minimality. For the benchmarks we considered in our evaluation, the strategy \emph{disjoint qubits} always generates results with minimum cost, whereas the strategy \emph{qubit triangle} yields the poorest results regarding minimality. Still, all strategies provide alternatives delivering solutions that are close to the minimum within acceptable runtime.

In a second series of evaluations, we compared the obtained minimal and close-to-minimal solutions to the mapping algorithm provided by IBM's Qiskit~\cite{qiskit}. This allows to evaluate how far existing heuristic approaches are from the optimum. To this end, we utilized the mapper available in Qiskit 0.4.15 and list the number of gates in the obtained circuit in the last column of Table~\ref{tab:results} (we do not list the runtime since all mappings could be determined within a second).\footnote{Note that, to ensure a fair comparison, we only considered the actual mapping process of Qiskit and not the decomposition as well as pre- or post-mapping optimizations that may be applied before/after the mapping.}
Since the mapping algorithm in Qiskit is probabilistic, we ran it 5 times for each benchmark and list the observed minimum.

As can be seen, the heuristic approach utilized in IBM's SDK Qiskit can be improved significantly. Considering the benchmarks \emph{alu-v3\_35} and \emph{mod5d1\_63
}, IBM's algorithm is 89\% and 104\% above the practical minimum that can be reached, respectively. On average, IBM's solution yields circuits that are 45\% above the minimum (by means of gate count).
Considering only the number of gates added during the mapping (and not the complete mapped circuit), Qiskit's solutions are 104\% above the minimum given by $\mathcal{F}$ on average---doubling the overhead required for mapping a circuit.
Hence, even though the exact approach proposed in this paper is only applicable for mapping small quantum circuits, it shows that there is much room for improvement of heuristic approaches---further motivating research on this topic.

\section{Conclusions}
\label{sec:conclusions}

In this paper, we proposed an exact solution for the mapping of quantum circuits to IBM QX architectures---an $\mathcal{NP}$-complete problem.
To this end, we  formulated the considered problem symbolically and used the Boolean satisfiability solver Z3 to determine a minimal solution. 
Moreover, we have shown how to improve the performance by restricting the search space while still guaranteeing close-to-minimal solutions.
By this, we do not only provide a method that maps quantum circuits to IBM's QX architectures with a \emph{minimal} number of SWAP and H operations, but also show by experimental evaluation that the number of operations added by IBM's heuristic solution exceeds the lower bound by more than 100\% on average. An implementation of the proposed methodology is publicly available
at \url{http://iic.jku.at/eda/research/ibm_qx_mapping}.

\begin{acks}
	This work has partially been supported by the LIT Secure and Correct System Lab funded by the State of Upper Austria and the European Union through the
	COST Action IC1405.
\end{acks}

\bibliographystyle{abbrv}
{
	\bibliography{../../bib/lit_header,../../bib/lit_mymisc,../../bib/lit_myrev,../../bib/lit_others,../../bib/lit_othersrev,../../bib/lit_rev,../../bib/lit_testing,../../bib/lit_misc,../../bib/lit_new,../../bib/lit_quantum,../../bib/lit_simulation,../../bib/lit_memristor}
}

\end{document}